\documentclass[
    ,final            % use final for the camera ready runs
  ]
  {aipproc}

\layoutstyle{8x11double}

\begin{document}

\title{Sub-mm/mm studies of the molecular gas in the Galactic disk; the TeV gamma ray SNR RXJ1713.7-3946 and the W28 high mass star forming region}


\author{Yasuo Fukui \& the NANTEN team}{
  address={Nagoya University, Chikusa-ku, Furocho, Nagoya 464-8602, Japan}
}

\begin{abstract}
Interstellar molecular clouds are sources of gamma rays through the interactions with cosmic ray protons followed by production of neutral pions which decay into gamma rays. We present new observations of the TeV gamma ray SNR RXJ 1713.7-3946 and the W28 region with the NANTEN2 sub-mm telescope in the $^{12}$CO $J=$2--1, 4--3 and 7--6 emission lines. 
In RXJ 1713.7-3946 we confirm that the local molecular gas having velocities around $-10$ km s$^{-1}$ (with respect to the local standard of rest) shows remarkably good spatial correlations with the SNR. We show that the X ray peaks are well correlated with the molecular gas over the whole SNR and suggest that the interactions between the SNR and the molecular gas play an important role in cosmic ray acceleration via several ways including magnetic field compression.
The $J=$4--3 distribution shows a compact and dense cloud core having a size of $\sim$1 pc as well as a broad wing towards peak C of the CO distribution. The cloud core shows a remarkable anti-correlation with the Suzaku X ray image and is also associated with hard gamma rays as observed with HESS. Based on these findings we present a picture that peak C is a molecular clump survived against the impact of the SN blast waves and is surrounded by high energy electrons emitting the X ray. The TeV gamma ray distribution is, on the other hand, more extended into the molecular gas, supporting the hadronic origin of gamma ray production. 
W 28 is one of the most outstanding star forming regions exhibiting TeV gamma rays as identified through a comparison between the NANTEN CO dataset and HESS gamma ray sources in the Galactic plane. In the W 28 region, we show the CO $J=$2--1 distribution over the whole region as well as the detailed image of the two TeV gamma ray peaks. One of the two peaks show strong CO $J=$7--6 emission, suggesting high excitation conditions in this high mass star forming core. A pursuit for the detailed mechanism to produce gamma rays is in progress. Finally, we discuss results of numerical simulation of the time dependent gamma ray image towards an ensemble of molecular clouds in the Galactic plane. 
\end{abstract}

\maketitle

%%%%%%%%%%%%%%%%%%%%%%%%%%%%%%%%%%%%%%%%%%%%
%% MAINMATTER
%%%%%%%%%%%%%%%%%%%%%%%%%%%%%%%%%%%%%%%%%%%%

\section{Introduction}

The interstellar molecular clouds in the Galaxy are gamma ray sources via the interaction between cosmic ray protons and the molecular gas which leads to production of neutral pions followed by decay into gamma rays. It is therefore important to understand the distribution of the molecular clouds in the Galactic disk and CO surveys have been used to identify such gamma ray sources extensively (e.g., Slane et al.\ 1999; Fukui et al.\ 2003).

RXJ 1713.7-3946 or G347.3-0.5 is a most remarkable SNR which exhibits non-thermal X rays and strong TeV gamma rays. This SNR was first discovered by ROSAT All Sky Survey (Pfeffermann \& Aschenbach 1996 ) and ASCA revealed that the X ray emission is dominated by synchrotron emission (Koyama et al.\ 1997; Slane et al.\ 1999). Subsequently to the discovery of TeV gamma rays by Cangaroo (Muraishi et al.\ 2000), HESS collaborations revealed the remarkable shell shape (Aharonian et al.\ 2006). RXJ 1713.7-3946 seems to be the most important object to understand the TeV gamma rays of hadronic origin because of its good association with molecular gas (Fukui et al.\ 2003; Moriguchi et al.\ 2005).

It has been shown that the W 28 region consisting of a SNR and high mass star forming regions show excellent coincidence with TeV gamma ray sources in the Galactic plane (Aharonian et al.\ 2008). This offers a new possibility that active star forming regions are the site of cosmic ray acceleration possibly via stellar winds, supernova explosions etc. of massive stars. 

We shall discuss the SNR and the star forming region in connection with the molecular gas revealed by new observations at sub-mm and mm wavelengths obtained with the NANTEN2 4m sub-mm telescope, which is an upgraded telescope of NANTEN, installed at an altitude of 4800 m in Atacama, Chile.

\begin{figure}
  \includegraphics[height=.4\textheight]{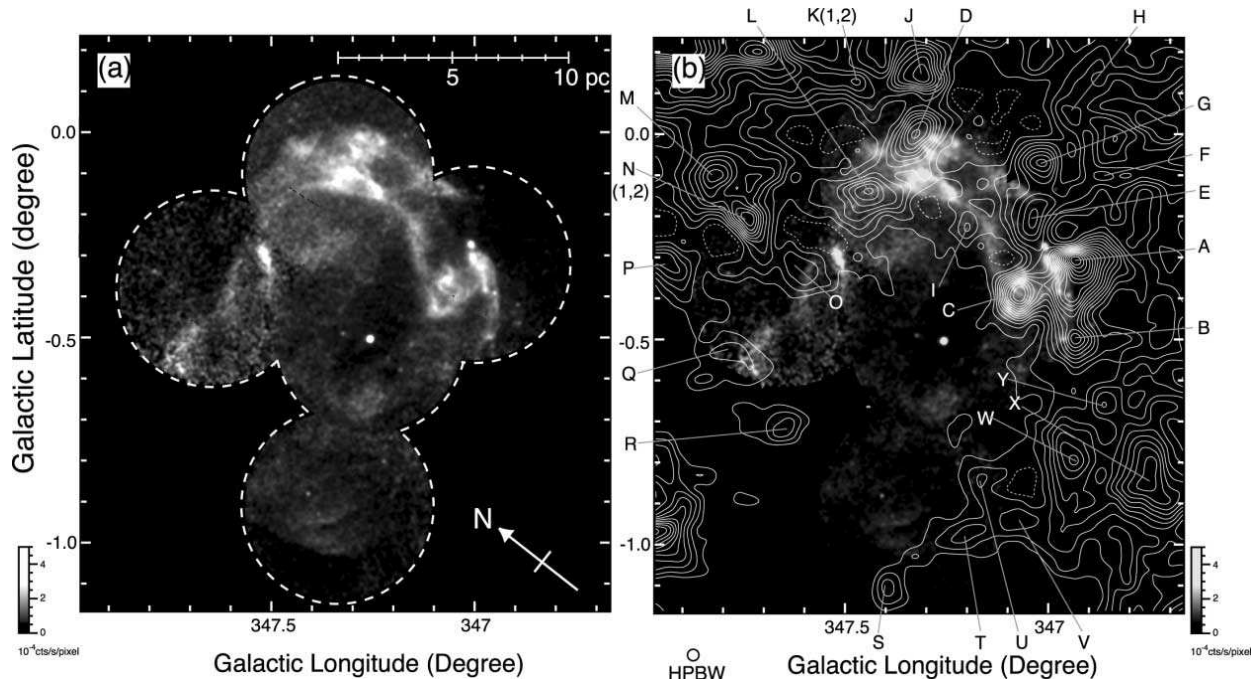}
  \caption{(a) X-ray image of G347.3-0.5 taken by XMM-Newton (Hiraga et al.\ 2005) in gray scale. An observed area is indicated with dashed lines.
(b) Overlay map of G347.3-0.5 XMM-Newton image in (a) and $^{12}$CO ($J=$1--0) intensity contours. The intensity is derived by integrating the $^{12}$CO ($J=$1--0) spectra from $-12$ to $-3$ km s$^{-1}$.
The lowest contour level and interval are 2.5 K km s$^{-1}$.
The depressions of the CO emission with the lowest contour level inside of the molecular boundary are shown with dashed contours. The CO peaks A-Y are indicated.
}
\end{figure}

\section{Distance of RXJ1713.7-3946}

The distance of the SNR has been under some debates. Koyama et al. (1997) estimated the distance to be 1 kpc from the X ray absorption column density.  This method assumes that the foreground HI in the line of sight is distributed uniformly and is subject to a large uncertainty if any localized cloud is responsible for the absorption. Slane et al.\ (1999) used the CO dataset obtained at 8.8 arcmin beam and identified that a CO cloud (called Cloud A) at $\sim$6 kpc is associated with part of the SNR. After that, 6 kpc was used as the distance widely in the literature. 

Fukui et al.\ (2003) made CO $J=$1--0 observations with a higher resolution of 2.6' beam with the NANTEN 4m telescope towards the SNR and inspected a large velocity range over 100 km s$^{-1}$ to search for any correlated features with the SNR. As a result, Fukui et al.\ (2003) discovered molecular gas at around $-10$ km s$^{-1}$ in $v_{LSR}$ shows the best correlation with the ASCA X ray image and concluded that the distance is $\sim$1 kpc based on the Doppler shift by the Galactic rotation. Koo (2004) made an independent analysis of HI and derived a similar distance of $\sim$1 kpc. A detailed account of the NANTEN observations were presented by Moriguchi et al.\ (2005). These authors confirmed the distance 1 kpc and noted that a foreground cloud perhaps located at a few 100 pc causes absorption in a soft X ray image which gives a constraint on the lower limit of the distance. Generally speaking, the random motion of the galactic clouds is around 5 km s$^{-1}$ and the error limits of 1 kpc are $\pm 0.5$ kpc from the kinematic argument.
 
Wang et al.\ (1997) proposed that a guest star recorded as AD 393 corresponds to the SNR. If we accept this identification, the age of the SNR is 1600 yr and the radius is estimated to be 8 pc for a nominal expansion velocity of 5000 km s$^{-1}$. We find that a distance 1 kpc is just consistent with this radius. Considering all these, we adopt the distance of the SNR to be 1 kpc.

\section{Correlations of molecular gas with the X rays and gamma rays}

It is important to note that the associated molecular gas can provide clues to understand the emission mechanisms of X rays and gamma rays. Figure 1 shows an overlay of the CO distribution on the XMM X ray image; the major four peaks of CO, A--D, (Fukui et al.\ 2003) and several additional CO peaks in Moriguchi et al.\ (2005) show remarkable correlations with the X ray distribution. We recognize that the X ray peaks are all associated with the CO peaks. Generally, the association is not exact positional matching but some shift of several arcmin is commonly seen; one of such examples is found towards peak A which is associated with a X ray peak towards ({\it l}, {\it b})$=$($346.933^{\circ}$, $-0.300^{\circ}$). Such association for all the X ray peaks strongly suggests that the X ray emitting components have physical connection with the molecular gas. 

Figure 2 shows an overlay of the NANTEN CO distribution with the TeV gamma ray image of HESS over the whole energy range (Aharonian et al.\ 2006). The CO distribution indicates a general agreement with the northwestern rim of the gamma ray distribution as discussed in detail by Aharoninan et al.\ (2006). Figures 3a and 3b, on the other hand, show overlays with the low- and high-energy bands of the HESS image. We find that peak C is associated with the strongest peak of the high-energy gamma rays.

\begin{figure}
  \includegraphics[height=.3\textheight]{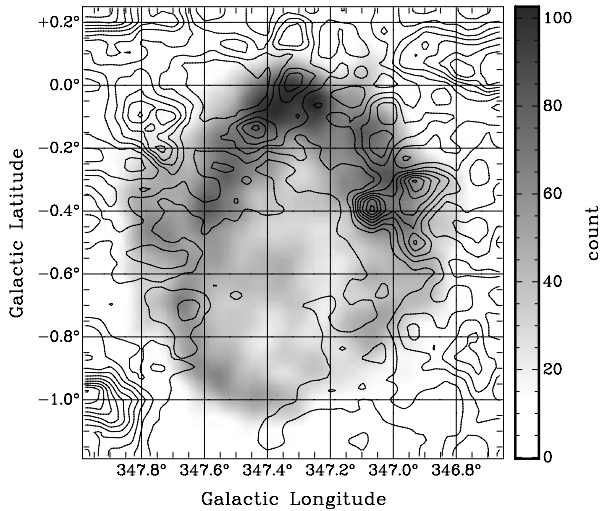}
  \caption{Gamma-ray image of G347.3-0.5 taken by HESS (Aharonian et al.\ 2007) in gray scale. Overlay map of G347.3-0.5 HESS image and $^{12}$CO ($J=$1--0) intensity contours. The intensity is derived by integrating the $^{12}$CO ($J=$1--0) spectra from $-18$ to 0 km s$^{-1}$. The lowest contour level and interval are 3 K km s$^{-1}$ and 5 K km s$^{-1}$, respectively.}
\end{figure}

\begin{figure}
  \includegraphics[height=.30\textheight]{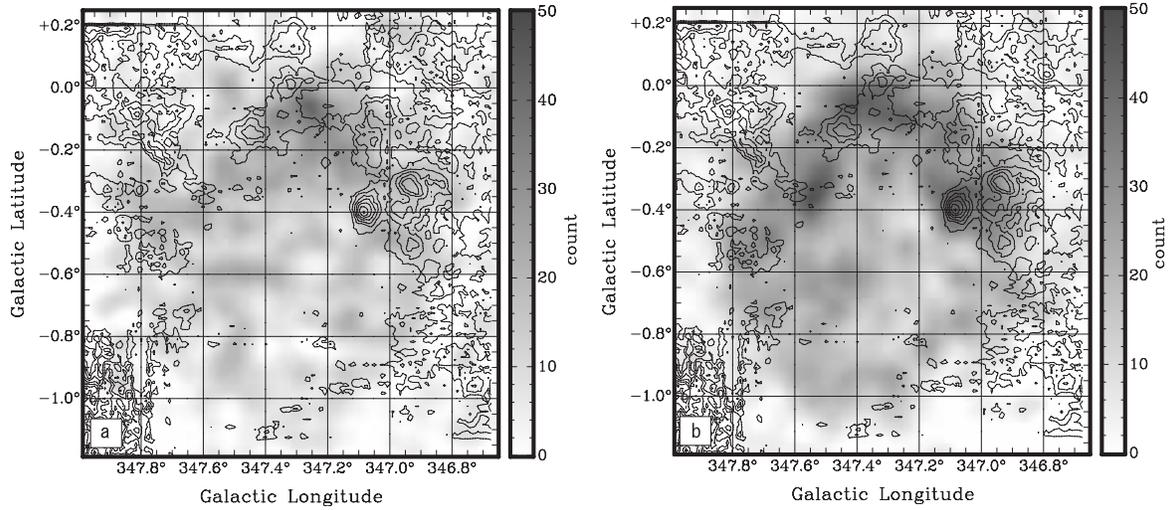}
  \caption{(a) Low-energy gamma-ray image of G347.3-0.5 taken by HESS (Aharonian et al. 2007) in gray scale. Overlay map of G347.3-0.5 HESS image and $^{12}$CO ($J=$2--1) intensity contours. The intensity is derived by integrating the $^{12}$CO ($J=$2--1) spectra from $-18$ to 0 km s$^{-1}$.
The lowest contour level and interval are 5.5 K km s$^{-1}$ and 5 K km s$^{-1}$, respectively.
(b)High-energy gamma-ray image of G347.3-0.5 taken by HESS (Aharonian et al.\ 2007) in gray scale.
The CO contours and the velocity range are the same as those in (a).}
\end{figure}

\begin{figure}
  \includegraphics[height=.30\textheight]{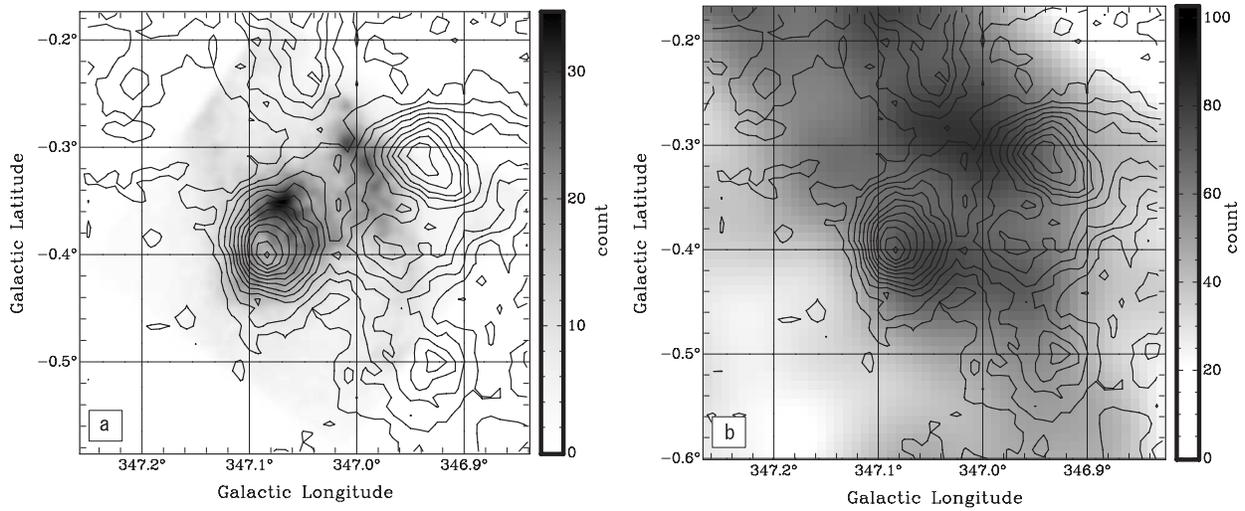}
  \caption{(a) X-ray image of peaks A, B, and C of G347.3-0.5 taken by Suzaku (Tanaka et al. 2008) in gray scale.
Overlay map of G347.3-0.5 Suzaku image and $^{12}$CO ($J=$2--1) intensity contours taken by NANTEN2. The intensity is derived by integrating the $^{12}$CO ($J=$2--1) spectra from $-16$ to $-3$ km s$^{-1}$.
The lowest contour level and interval are 3 K km s$^{-1}$ and 2.75 K km s$^{-1}$, respectively.
(b) Gamma-ray image of peaks A, B, and C of G347.3-0.5 taken by HESS (Aharonian et al.\ 2007) in gray scale.
Overlay map of G347.3-0.5 HESS image and $^{12}$CO ($J=$2--1) intensity contours taken by NANTEN2.
The CO contours and the velocity range are the same as those in (a).}
\end{figure}

The TeV gamma image is generally taken to indicate a shell shape similar to the X ray images (e.g., Tanaka et al.\ 2008) but it is not true that the gamma ray and X ray show exact agreement with each other. The HESS image in fact shows some significant differences with the X ray image on smaller scales, for instance, towards peak C (see Figure 4).

\section{Star formation around RXJ1713.7-3946}

It has been suggested that the supernova explosion occurred within the wind-blown bubble created by the massive star projenitor star (Slane et al. 1999 and Ellison et al. 2001). This scenario is consistent with the candidate for a central neutron star which resulted from a core-collapse supernova as suggested by Slane et al. (1999); Lazendic et al. (2003). This bubble is likely forming a shell of molecular gas currently interacting with the SNR. Such a wind-blown bubble is observed elsewhere in the CO emission (e.g., Yamamoto et al. 2004). 

If we adopt the timescale of the bubble to be $\sim $10Myrs, the compressed gas may be able to form stars in it. This scenario is in fact consistent with the identification of the young stars based on the IRAS point source catalog (Moriguchi et al. 2005). These authors note three IRAS point sources are located towards peaks A, B and C, which are low to intermediate-mass protostars of 100-1000 solar luminosities with protostellar far-infrared SED. Accordingly, we are witnessing an ad-hoc situation where the supernova blast waves just reached the molecular shell and initiated the interaction whose duration may be only a few 100yrs at most.

\section{Peaks A and C; NANTEN2 sub-mm results}

Sub-mm line emission of interstellar molecules is an excellent probe of the dense and warm molecular gas. Figures 4a and 4b show the distribution of the $^{12}$CO $J=$2--1 emission line obtained with the NANTEN2 sub-mm telescope overlayed on the Suzaku X ray and HESS TeV gamma ray distributions, respectively. The $^{12}$CO $J=$4--3 distribution in Figures 5a and 5b indicates a dense and compact core smaller than 1pc whose density is around 10$^4$ cm$^{-3}$ or higher overlayed on the same images of X ray and gamma ray. The core shows a good agreement with the IRAS point source IRAS17089-3951 and the broad wings first discovered by Fukui et al. (2003).

\begin{figure}
  \includegraphics[height=.30\textheight]{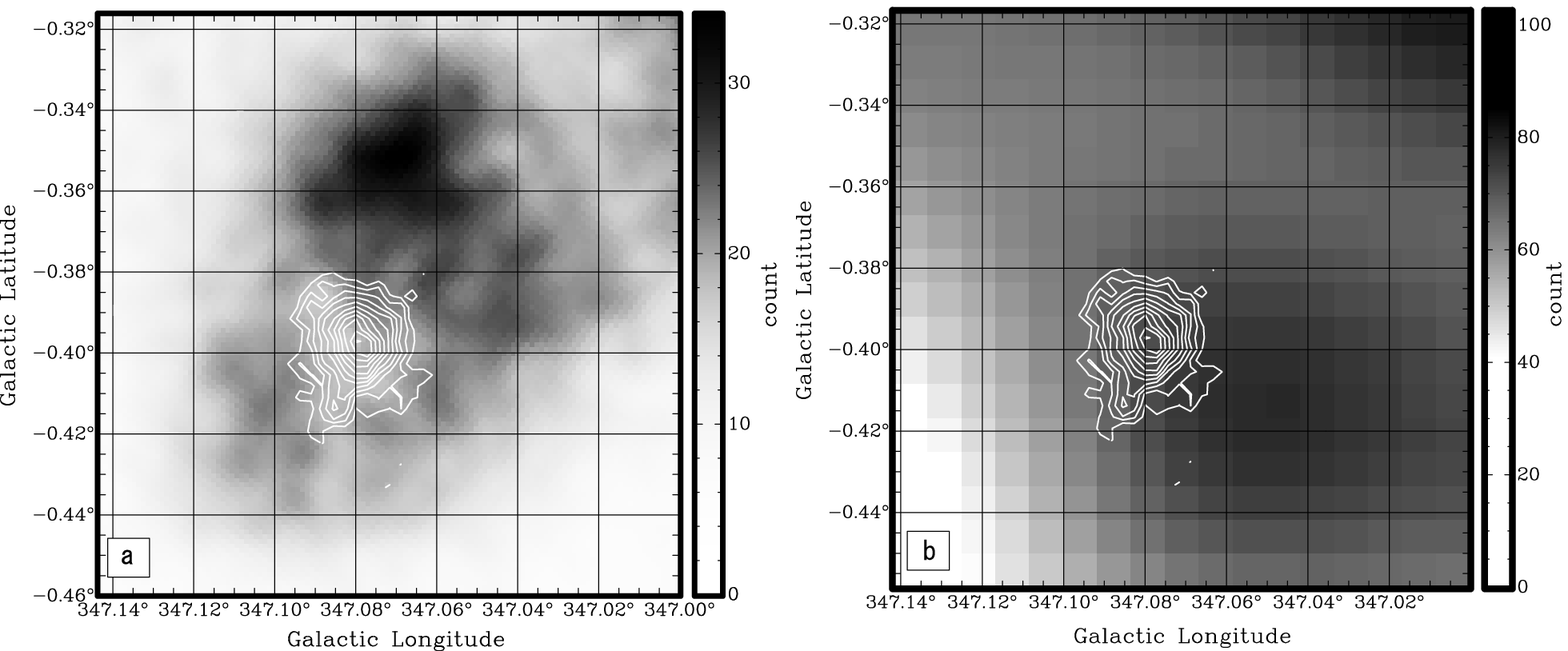}
  \caption{(a) X-ray image of peaks A, B, and C of G347.3-0.5 taken by Suzaku (Tanaka et al.\ 2008) in gray scale.
Overlay map of G347.3-0.5 Suzaku image and $^{12}$CO ($J=$4--3) intensity contours taken by NANTEN2. The intensity is derived by integrating the $^{12}$CO ($J=$4--3) spectra from $-20$ to $-4$ km s$^{-1}$.
The lowest contour level and interval are 10 K km s$^{-1}$ and 2.5 K km s$^{-1}$, respectively.
(b) Gamma-ray image of peaks A, B, and C of G347.3-0.5 taken by HESS (Aharonian et al.\ 2007) in gray scale.
Overlay map of G347.3-0.5 HESS image and $^{12}$CO ($J=$4--3) intensity contours taken by NANTEN2.
The CO contours and the velocity range are the same as those in (a).}
\end{figure}

\begin{figure}
  \includegraphics[height=.30\textheight]{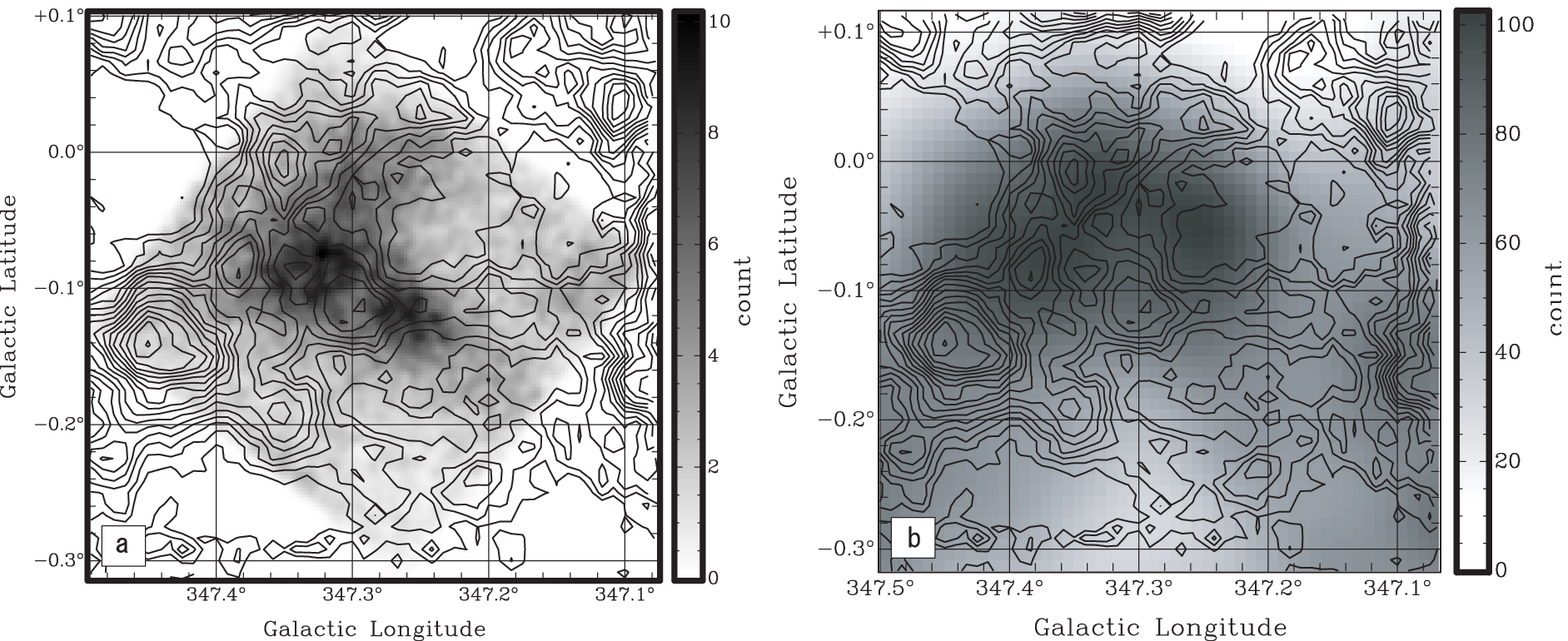}
  \caption{(a) X-ray image of peak D of G347.3-0.5 taken by Suzaku (Tanaka et al.\ 2008) in gray scale.
Overlay map of G347.3-0.5 Suzaku image and $^{12}$CO ($J=$2--1) intensity contours taken by NANTEN2.The intensity is derived by integrating the $^{12}$CO ($J=$2--1) spectra from $-16$ to $-10$ km s$^{-1}$.
The lowest contour level and interval are 1.5 K km s$^{-1}$ and 0.75 K km s$^{-1}$, respectively.(b) Gamma-ray image of peak D of G347.3-0.5 taken by HESS (Aharonian et al.\ 2007) in gray scale.
Overlay map of G347.3-0.5 HESS image and $^{12}$CO ($J=$2--1) intensity contours taken by NANTEN2.
The CO contours and the velocity range are the same as those in (a).}
\end{figure}

These agreements indicate that the IRAS source is an embedded young low-mass protostar driving the molecular outflow instead of shock-accelerated wings suggested before (Fukui et al.\ 2003; Moriguchi et al.\ 2005). We suggest that the $J=$4--3 core represents a dense core embedded in the cavity swept-up by the blast waves of the supernova remnant. We note that peak C is surrounded by thin X ray features which is likely emitted from the volume close to the surface of the core (Figure 5). In addition, peak A is well delineated by the X ray emission, suggesting possibly a nearly edge on view of the impacting blast waves on the molecular gas (Figure 4). The fact that peak C is most prominent in the high-energy band of HESS image suggests that the cosmic ray protons are actively produced in its vicinity. Further, the $^{12}$CO $J=$2--1 distribution of peak D (Figures 6a  and 6b) shows also similar anti-correlation with the Suzaku X ray image, indicating that the situation discussed for peak C seems commonly applicable to peak D.

\begin{table}
\begin{tabular}{lccc}
\hline
  & \tablehead{1}{r}{b}{RXJ1713.7-3946\tablenote{Moriguchi et al.\ (2005)}}
  & \tablehead{1}{r}{b}{Vela Jr.\tablenote{Aharonian et al.\ (2007)}}
  & \tablehead{1}{r}{b}{RCW 86\tablenote{Rosad et al.\ (1996)}}\\
\hline
age (yr) & 1600 & 500--5000 & 9500 \\
mass ($M_{\odot}$) & 5500 & 1.4--14 & 4100 \\
distance (kpc) & 1.0 &  0.2--1.0 & 2.8 \\
\hline
\end{tabular}
\caption{List of age, mass and distance of each object.}
\label{tab:a}
\end{table}

\section{Comparison with the other TeV gamma ray SNRs}

We shall here briefly compare RXJ 1713.7-3946 with the other SNRs, Vela Jr., and RCW 86. We have studied the CO distribution with NANTEN towards these SNRs and list mass of the associated molecular gas in Table 1. It is notable that RXJ 1713.7-3946 is associated with quite massive molecular gas for its young age. By considering their shape, we suggest that the expansion of RXJ 1713.7-3946 is most significantly affected by the surrounding molecular gas although the others are much less so at the present epoch. This is consistent with the circular shape of the others as compared with RXJ 1713.7-3946 whose shape is distorted into a parallelogram-like shape (e.g., Uchiyama et al.\ 2005). So, RXJ 1713.7-3946 is unique in its interaction with the massive molecular gas that is distributed over three quarters of its outer boundary (Moriguchi et al.\ 2005).

The interaction of the SNR has been discussed in detail from a view point of gamma$/$X rays by Aharonian et al.\ (2006) and Uchiyama et al.\ (2005). A recent analysis of the Suzaku and HESS data, on the other hand, did not explicitly discuss the connection with the molecular gas (Tanaka et al.\ 2008). Based on a spectral analysis these authors concluded that the gamma rays are produced by the hadronic process with magnetic field strength of $\sim 100 \mu G$ rather than the leptonic process because the latter cannot reproduce the low-energy TeV spectrum without introducing some ad-hoc parameters of background photons. Also, they noted that the distance of the SNR may be closer than 1kpc to explain the cosmic ray proton energy in a reasonable range less than $10^{51}$ ergs under the assumption that the target proton density is $\sim $1 cm$^{-3}$. 

The present argument given above indicates that the molecular gas plays an important role in the emission properties of the gamma$/$X rays. We shall note that the field strength adopted by Tanaka et al.\ (2008) may actually be a result of the expansion of the SNR into the dense gas and the effects of the interaction are implicitly taken into account in the field strength. Since X rays are emitted in the lower density volume adjacent to the molecular gas where density is low, the molecular gas only indirectly affects the X ray properties except for the magnetic field strength and perhaps for locally enhanced density. For the hadronic gamma ray production, the total cosmic ray proton energy is expressed as Wp $\sim 2 \times 10^{50}$ ($n/1$ cm$^{-3}$)$^{-1}$($d/$1 kpc)$^{2}$ [erg] (Tanaka et at.\ 2008) and a distance smaller than 1 kpc is favored if density is taken to be 1 cm$^{-3}$. We suggest that the close connection between the HESS gamma ray and the NANTEN molecular gas suggests that the typical interacting density may be higher than 1 cm$^{-3}$ since the molecular gas has density of $\sim $100 cm$^{-3}$ or more and that a smaller distance is not required.

The current theoretical studies of the cosmic ray acceleration are based on diffusive shock acceleration (for review, see Blandford and Eichler 1987; Malkov and Drury 2001). We suggest that inclusion of the interaction with the molecular gas may remain to be more explicitly made in such theories for better understanding of the emission properties of the gamma$/$X rays.

\begin{figure}
  \includegraphics[height=.35\textheight]{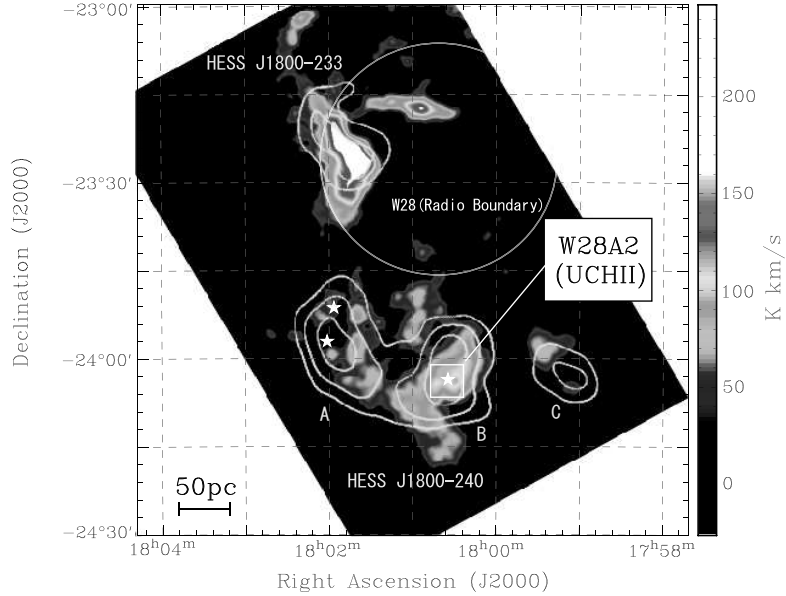}
  \caption{NANTEN2 $^{12}$CO ($J=$2--1) image of the W 28 region for $V_{\rm LSR} = -10$ to 25 km s$^{-1}$ with HESS gamma-ray significance contours overlaid at levels of 4, 5, and 6 $\sigma$. The location of the Hll region (white stars) are indicated (Aharonian et al.\ 2008). As for the region enclosed by a square, more detailed observations were done in the $^{12}$CO ($J=$4--3) and $^{12}$CO ($J=$7--6) emissions with the NANTEN2 telescope.}
\end{figure}
\begin{figure}
  \includegraphics[height=.30\textheight]{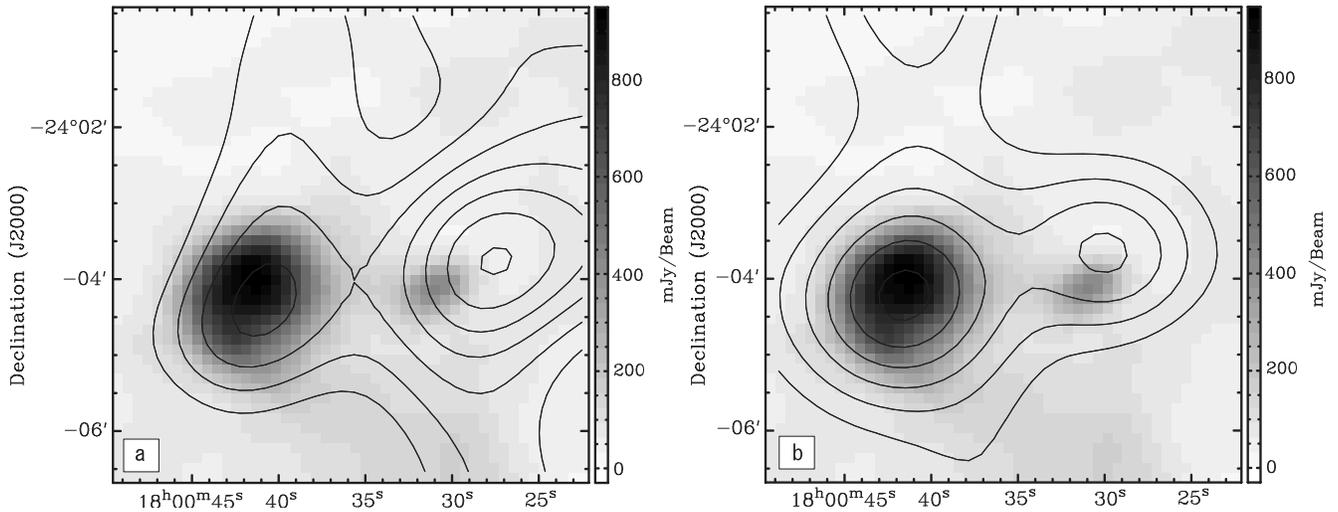}
  \caption{(a) The distribution of $^{12}$CO ($J=$4--3) integrated intensity obtained with NANTEN2 overlayed on a 20 cm (VLA image) map.
The velocity integration range is from $-10$ km s$^{-1}$ to 25 km s$^{-1}$. The lowest contour level is 50 K km s$^{-1}$ and the contour interval is 5 K km s$^{-1}$ with a 106'' beam.
(b) The contours indicate the $^{12}$CO ($J=$7--6) integrated intensity
obtained with NANTEN2. The lowest contour level is 20 K km s$^{-1}$ and the contour interval is 5 K km s$^{-1}$.}
\end{figure}

\section{High mass star forming region W 28}

W 28 is one of the most outstanding star forming regions exhibiting TeV gamma rays as identified through a comparison between the NANTEN CO dataset and HESS gamma ray sources in the Galactic plane. It is most outstanding in the gamma ray-CO correlation among the similar objects discovered so far. Other sources of high mass star forming regions like Westerlund2 shall be discussed elsewhere.

Figure 7 shows the large scale $^{12}$CO ($J=$2--1 distribution overlayed with the HESS TeV gamma rays. This clearly indicates a good spatial correlation between them. It is not yet conclusive but is possible that all the clouds in Figure 7 are physically connected with each other since their LSR velocities are in a range of $\sim 20$ km s$^{-1}$; an association of giant molecular clouds is possible at a distance of a few kpc (Aharonian et al.\ 2008 and references therein).

Figure 8 shows the two cloud cores in the south of the W28 region as observed in the $J=$7--6 and 4--3 emission lines with NANTEN2. These high excitation lines indicate that density may be as large as $10^{5}$cm$^{-3}$ in the centre of these cores which results in very active formation of high mass stars associated with high velocity molecular outflow. Further follow-up studies of these cores at various wavelengths should clarify more details relevant to the gamma ray production.

Recent numerical simulations which takes into account the actual molecular distributions obtained with NANTEN by assuming some time dependent SNe (Casanova et al.\ 2008) offers a new insight into a 10-degree-scale gamma-ray view of the Galactic plane and will provide a better insight into the sensitive gamma ray data becoming available soon with Fermi, HESS, etc.

\section{Summary}
We presented the new molecular data of RXJ 1713.7-3946 and compared them with the Suzaku X ray and HESS gamma ray distributions as well as those in the W28 star forming region. The main conclusions are summarized below.

\begin{enumerate}
\item We have strong indications that the molecular gas is interacting with RXJ1713.7-3946 extensively over its entire spatial extent. This interaction is demonstrated by the close association of molecular gas with the X ray peaks as well as the large-scale correlation of the SNR shell of the gamma$/$X rays with the associated molecular gas. We have several reasons to adopt the distance of RXJ1713.7-3946 to be $\sim$1 kpc.

\item The molecular peaks A, C and D exhibit particularly strong correlations with the X rays. Based on the new sub-mm CO data we argue that peak C is a pre-existent dense core currently forming a protostar. This core survived against the impact of the SNR blast waves thanks to its high density of$\sim$$10^4$ cm$^{-3}$ or more and is embedded within the SNR. It is interesting to note that this unique location may favor the higher gamma ray production as indicated by the very hard gamma ray spectrum towards peak C.

\item The distributions of the HESS gamma rays show differences from that of the X rays on smaller scales of $\sim$1 pc or less. We suggest that this is caused by that the X rays are mainly emitted from the volume near the surface of the molecular gas while the gamma rays from the interiors of the molecular gas where cosmic rays protons penetrated. In order to prove this suggestion, it is highly desirable to develop theories of particle acceleration with a strong density gradient due to molecular gas.

\item The $^{12}$CO $J=$2--1 distribution obtained with NANTEN2 indicates a detailed correlation between TeV gamma rays and molecular gas in the W28 region. The $^{12}$CO $J=$4--3 and 7--6 images have further revealed the density is as high as $\sim 10^5$ cm$^{-3}$, supporting the youth of the dense cores. High mass young stars are one of the promising candidates for gamma ray production under the strong effects of stellar winds, SNe, etc.

\end{enumerate}

\begin{theacknowledgments}
The authors are grateful to the Scientific Organizing Committee of the Conference for the kind invitation to this exciting meeting.
NANTEN2 is an international collaboration among 10 universities including Universities of Cologne and Bonn.
We greatly appreciate the hospitality of all staff members
of the Las Campanas Observatory of the Carnegie Institution
of Washington. The NANTEN telescope was operated based
on a mutual agreement between Nagoya University and the
Carnegie Institution of Washington. We also acknowledge
that the operation of NANTEN can be realized by contributions
from many Japanese public donators and companies.
This work is financially supported in part by a Grant-in-
Aid for Scientific Research (KAKENHI) from the Ministry
of Education, Culture, Sports, Science and Technology of
Japan (Nos. 15071203 and 18026004) and from JSPS (Nos.
14102003, 20244014, and 18684003). This work is also financially
supported in part by core-to-core program of a Grantin-
Aid for Scientific Research from the Ministry of Education,
Culture, Sports, Science and Technology of Japan (No. 17004).
\end{theacknowledgments}


\begin{thebibliography}{9}
%\bibitem{Slane1997}
%P.~Slane, \emph{AAS} \textbf{192}, 9904S (1997)

\bibitem{Slane1999}
P.~Slane, \emph{ApJ} \textbf{525}, 357 (1999)

\bibitem{Fukui2003}
Y.~Fukui, \emph{PASJ} \textbf{55}, L61 (2003)

\bibitem{Pfeffermann & Aschenbach1996}
E.~Pfeffermann, \& B.~Aschenbach, \emph{MPE Rep.} \textbf{263}, 267 (1996)

\bibitem{Koyama1997}
K.~Koyama, \emph{PASJ} \textbf{49}, L7 (1997)

\bibitem{Muraishi2000}
H.~Muraishi, \emph{A\&A} \textbf{354}, L57 (2000)

\bibitem{Aharonian2006}
F.~Aharonian, \emph{A\&A} \textbf{449}, 223 (2006)

\bibitem{Moriguchi2005}
Y.~Moriguchi, \emph{ApJ} \textbf{631}, 947 (2005)

\bibitem{Aharonian2008}
F.~Aharonian, \emph{A\&A} \textbf{481}, 401A (2008)

\bibitem{Koo2004}
B.~-C.~Koo, \emph{Korean Astron. Soc.} \textbf{37}, 61 (2004)

\bibitem{Wang1997}
Z.~R.~Wang, \emph{A\&A} \textbf{318}, L59 (1997)

\bibitem{Tanaka2008}
Y.~Moriguchi, \emph{ApJ} \textbf{685}, 988T (2008)

\bibitem{Ellison2001}
D.~C.~Ellison, \emph{ApJ} \textbf{563}, 191 (2001)

\bibitem{Lazendic2003}
J.~S.~Lazendic, \emph{ApJ} \textbf{593}, L27 (2003)

\bibitem{Yamamoto2005}
H.~Yamamoto, \emph{prpl.conf.} \textbf{8295Y} (2005)

\bibitem{Uchiyama2005}
Y.~Uchiyama, \emph{High Energy Gamma-Ray Astronomy} \textbf{745}, 305 (2005)

\bibitem{Blandford & Eichler}
R.~Blandford, \& D.~Eichler, \emph{1987PhR} \textbf{154}, 1B (1987)

\bibitem{Malkov & Drury}
M.~A.~Malkov, \& 0'C.~Drury, \emph{2001RPPh} \textbf{64}, 429M (2001)

\bibitem{Casanova2008}
S.~Casanova, \emph{Astro-ph} \textbf{arXiv:0810.4297} (2008)

\bibitem{Hiraga2005}
J.~S.~Hiraga, \emph{A\&A} \textbf{431}, 953 (2005)

\bibitem{Aharonian2007}
F.~Aharonian, \emph{ApJ} \textbf{661}, 236 (2007)

\bibitem{Rosad1996}
M.~Rosad, \emph{A\&A} \textbf{315}, 243 (1996)
\end{thebibliography}
\end{document}